\documentclass[twocolumn,aps,showpacs,superscriptaddress,aps]{revtex4}
\usepackage{mathrsfs}
\usepackage{amsfonts}
\usepackage{graphicx}
\usepackage[latin1]{inputenc}
\usepackage{amsmath}
\usepackage{amsfonts}
\usepackage{amssymb}
\usepackage{amsthm}
\usepackage{amsxtra}
\usepackage{fancyhdr}
\usepackage{mathrsfs}
\usepackage{amsmath,amssymb,amsmath}
\usepackage{graphicx}% Include figure files
\usepackage{dcolumn}% Align table columns on decimal point
\usepackage{bm}% bold math
\begin{document}
\title{Complexity analysis of Klein-Gordon single-particle systems}

\author{D. Manzano}
\email{manzano@ugr.es}
\address{Departamento de F\'isica At\'omica, Molecular y Nuclear and 
Instituto Carlos I de F\'isica Te\'orica y Computacional, 
Universidad de Granada, 18071-Granada,
Spain}

\author{S. L\'opez-Rosa}
\email{slopez@ugr.es}
\address{Departamento de F\'isica At\'omica, Molecular y Nuclear and 
Instituto Carlos I de F\'isica Te\'orica y Computacional, 
Universidad de Granada, 18071-Granada,
Spain}

\author{J.S. Dehesa}
\email{dehesa@ugr.es}
\address{Departamento de F\'isica At\'omica, Molecular y Nuclear and 
Instituto Carlos I de F\'isica Te\'orica y Computacional, 
Universidad de Granada, 18071-Granada,
Spain}
% \address{Instituto Carlos I de F\'isica Te\'orica y Computacional, Universidad de Granada,
% 18071-Granada, Spain}


\begin{abstract}
 The Fisher-Shannon complexity is used to quantitatively estimate the contribution of
 relativistic effects to on the internal disorder of Klein-Gordon single-particle Coulomb systems 
 which is manifest in the rich variety of three-dimensional geometries of its corresponding
 quantum-mechanical probability density. It is observed that, contrary to the non-relativistic case, the Fisher-Shannon 
 complexity of these relativistic systems does depend on the potential strength (nuclear charge). 
 This is numerically illustrated for pionic atoms. Moreover, its variation with the quantum numbers
 $(n,l,m)$ is analysed in various ground and excited states. It is found that the relativistic 
 effects enhance when $n$ and/or $l$ are decreasing.
\end{abstract}

\maketitle
Numerous phenomena and properties of many-electron systems have been qualitatively characterized by information-theoretic
 means. In particular, various single and composite information-theoretic measures have been proposed to identify and analyze 
the multiple facets of the internal disorder of non-relativistic quantum systems; see e.g. Ref. \cite{G09,L95,S99,PIP,R04,D09,A08a,A09,P07}.

They are often expressed as products of two quantities of local (e. g. the Fisher information) and/or 
global (e. g. the variance or Heisenberg measure, the Shannon entropy, the Renyi and Tsallis entropies 
and the disequilibrium or linear entropy $\left<\rho\right>$) character, which describe the charge 
spreading of the system in a complementary and more complete manner than their individual components. 
This is the case of the disequilibrium-Shannon or Lopez-Ruiz-Mancini-Calvet (LMC in short) \cite{L95}, 
disequilibrium-Heisenberg \cite{S99}, Fisher-Shannon \cite{R04,A08a} and the Cramer-Rao  \cite{C91,A08a} 
complexities, which have their minimal values at the extreme ordered and disordered limits.

Recently these studies have been extended to take into account the relativistic effects in atomic physics. 
Relativistic quantum mechanics \cite{G00} tells us that special relativity provokes (at times, severe) spatial 
modifications of the electron density of many-electron systems, what produces fundamental and measurable changes
in their physical properties. The qualitative and quantitative evaluation of the relativistic modification of 
the spatial redistribution of the electron density of ground and excited states in atomic and molecular systems 
by information-theoretic means is a widely open field. In the last three years the relativistic effects of various
single and composite information-theoretic quantities of the \underline{ground states} of hydrogenic \cite{K09} 
and neutral atoms \cite{B07,S08,S09}  have been investigated in different relativistic settings. 

First Borgoo et al \cite{B07} (see also \cite{B08}), in a Dirac-Fock setting, find that the LMC shape complexity 
of the ground-state atoms (i) has an increasing dependence on the nuclear charge (also observed by Katriel and Sen 
\cite{K09} in Dirac ground-state hydrogenic systems), (ii) manifest  shell and relativistic effects, the latter 
being specially relevant in the disequilibrium ingredient (which indicates that they are dominated by the innermost
orbital). Then, Sa\~nudo and L\'opez-Ruiz \cite{S09} (see also \cite{S08}) show a similar trend for both LMC and 
Fisher-Shannon complexities in a different setting which uses the fractional occupation probabilities of electrons
in atomic orbitals instead of the continuous electronic wavefunctions; so, they use discrete forms for the 
information-theoretic ingredients of the complexities. Moreover, their results allow to identify the shell structure 
of noble gases and the irregular shell filling of some specific elements; this phenomenon is specially striking in 
the Fisher-Shannon case as the authors explicitly point out. 

The present work contributes to this new field with the quantification of the relativistic compression of both ground 
and excited states of the Klein-Gordon single-particle wavefunctions in a Coulombian well by means of the Fisher-Gordon 
complexity. This quantity is defined by

\begin{equation}
C_{FS}\left[\rho\right]:= I \left[ \rho \right] \times  J \left[ \rho \right], \hspace*{1cm}
\label{eq:fs}
\end{equation}
where

\begin{equation}
I \left[ \rho \right] = \int \rho \left[ \frac{d}{dx} \ln \rho \left(\vec{r} \right) \right]^2   d\vec{r}, \; 
J  \left[ \rho \right]= \frac{1}{2 \pi e} \exp \left( 2 S \left[ \rho \right] /3 \right),
\end{equation}
are the Fisher information and the Shannon entropic power of the density $\rho(\vec{r})$, respectively. The latter quantity, 
which is an exponential function of the Shannon entropy $S[\rho]=- \left< ln \rho \right>$, measures the total extent to which 
the single-particle distribution is in fact concentrated \cite{C91}. The Fisher information $I\left[ \rho \right]$, which is 
closely related to the kinetic energy \cite{H09}, is a local information-theoretic quantity  because it is a gradient 
functional of the density, so being sensitive to the single-particle oscillations. Then, contrary to the remaining complexities
published in the literature up until now, the Fisher-Shannon complexity has a property of locality and it takes simultaneously
into account the spatial extent of the density and its (strong) oscillatory nature. 

Here we use the Fisher-Shannon complexity to quantify the relativistic charge spreading of Klein-Gordon particles moving in a 
Coulomb potential $V(\vec{r})=-\frac{Ze^2}{r}$. We study the dependence of these quantities on the potential strength $Z$ and 
on the quantum numbers $(n,l,m)$ which characterize the stationary states of a spinless relativistic particle with a negative 
electric charge. 

The Klein-Gordon wave equation was introduced in 1926 and constituted the first theoretical description of
 particle dynamics in a relativistic quantum setting \cite{S26}. Since then, the study of its properties for different potentials of various
 dimensionalities has been a problem of increasing interest \cite{N79, M84,C06,A08}. Many efforts were addressed to obtain the spectrum 
of energy levels and the ordinary moments or expectation values $\left< r^k\right>$ of the charge distribution of numerous single-particle 
systems (such as e.g. muonic and pionic atoms \cite{B79}). Only recently, Chen and  Dong \cite{C06} have been able to calculate explicit 
expressions for these moments and some off-diagonal matrix elements of $r^k$ for a Klein-Gordon single-particle of mass $m_0$ in the 
Coulomb potential $V(\vec{r})=-\frac{Ze^2}{r}$. These authors, however, do not use the Lorentz-Invariant (LI) Klein-Gordon charge density

 \begin{equation}\label{eq:LI}
\rho_{LI}(\vec{r})=\frac{e}{m_0 c^2}\left[ \epsilon-V(r)\right]  \left|\Psi_{n l m}(\vec{r}) \right|^2,
\end{equation}
but just the Non-Lorentz-Invariant (NLI) expression $\rho_{NLI}(\vec{r})=\left|\Psi_{n l m}(\vec{r}) \right|^2$ as in the 
non-relativistic case, where $\epsilon$ and $\Psi(\vec{r})$ denote the physical solutions of the Klein-Gordon equation \cite{N79,C06}

\begin{equation}\label{eq:KG}
\left[ \epsilon-V(r) \right] \psi(\vec r)=( - \hbar^2 c^2 \nabla^2+m_0^2c^4 ) \psi(\vec r),
\end{equation}
which characterizes the wavefunctions $\Psi_{n l m}(\vec{r},t)=\psi_{n l m}(\vec{r})exp\left( -\frac{i}{\hbar}\epsilon t \right)$ 
of the stationary states of our system. In spherical coordinates $\vec{r}=(r,\theta,\phi)$, the eigenfunction 
$\psi(r,\theta,\psi)=r^{-1}u(r)Y_{lm}(\theta,\phi)$, where the $Y$-symbol denotes the spherical harmonics of order $(l,m)$. 
Making the change $r\to s$, with $s=\beta r$ in Eq. (\ref{eq:KG}), and using the notations

\begin{equation}
\beta \equiv \frac{2}{\hbar c}\sqrt{m_0^2c^4-\epsilon^2};\qquad \lambda\equiv \frac{2\epsilon Z e^2}{\hbar^2 c^2 \beta},
\end{equation}
one has the radial Klein-Gordon equation 

\begin{equation}\label{eq:radial}
\frac{d^2 u(s)}{ds^2}-\left[ \frac{l'(l'+1)}{s^2}-\frac{\lambda}{s}+\frac{1}{4} \right] u(s)=0,
\end{equation}
where the following additional notations 

\begin{equation}
l'=\sqrt{\left( l+\frac{1}{2}  \right)^2-\gamma^2}-\frac{1}{2},\qquad with \quad \gamma \equiv Z \alpha,
\end{equation}
have also been used, being the fine structure constant $\alpha\equiv \frac{e^2}{\hbar c}$. It is known 
(see e.g. Eq. (21a) in page 39 of Ref \cite{G00} and Eq (5.12) with $N=1$ of Ref. 
\cite{N79} duly corrected for the inappropriate location of the $\frac{1}{2}$-power in it)
that the bound states have the energy eigenvalues 

\begin{equation}
\epsilon=\frac{m_0c^2}{\sqrt{1+\left( \frac{\gamma}{n-l+l'}\right)^2}}
\end{equation}
and the eigenfunctions 

\begin{equation}\label{eq:red}
u_{n l}(s)=\mathscr{N} s^{(l'+1)}e^{-\;\frac{s}{2}}\widetilde{L}_{n-l-1}^{2l'+1}(s).
\end{equation}

To preserve Lorentz invariance, according to relativistic quantum mechanics \cite{G00}, we calculate the constant 
$\mathscr{N}$ by taking into account the charge conservation $\int_{\mathbb{R}^3} \rho(\vec{r})d^2r=e$, which yields
the value \cite{M08} 

\begin{equation}\label{eq:N}
\mathscr{N}^2=\frac{m_0c^2\gamma}{\hbar c} \frac{1}{(n+l'-l)^2+\gamma^2},
\end{equation}

Let us emphasize that the resulting Lorentz-invariant charge density $\rho_{LI}(\vec{r})$ given by Eqs. (\ref{eq:LI})-(\ref{eq:N}) is always (i.e., 
for any observer's velocity $v$) appropriately normalized, while the non-Lorentz-invariant density $\rho_{NLI}(\vec{r})$ used by Cheng and Dong 
\cite{C06} is not. This was numerically discussed in Ref \cite{M08} for some pionic atoms in the infinite nuclear mass approximation.

% \begin{figure}
% \begin{center}
% \includegraphics[scale=0.5]{figure1.eps}
% \end{center}
% \caption{Normalization of the charge density for the Lorentz invariant (LI) and the non-Lorentz invariant (NLI) charge densities.}
% \end{figure}

Let us now numerically discuss the relativistic effects in the Fisher-Shannon complexity of a pionic system. First, 
we center our attention in the dependence on the nuclear charge of the system. As we can see in Figure 1, the 
Fisher-Shannon complexity of the Klein-Gordon case depend on the nuclear charge $Z$, contrary to the non-relativistic 
description. The Schr\"odinger or non-relativistic value of the Fisher-Shannon complexity 
has been recently shown to be independent of the nuclear charge $Z$ for any hydrogenic system, (\cite{SKA}, see also \cite{R04}).
  It is apparent that this quantity is a very good indicator of the relativistic effects as has been recently pointed out by Sa\~nudo and L\'opez-Ruiz
 \cite{S09,S08} in other relativistic settings. These effects are bigger when the nuclear charge increases, so the relativistic Fisher-Shannon complexity
 enhances. This behaviour is easy to understand because when we take into account the relativistic effects, the charge probability density is more
 compressed towards the nucleus than in the non-relativistic case \cite{G00,M08}.

\begin{figure}[h]
\begin{center}
\includegraphics[scale=0.45]{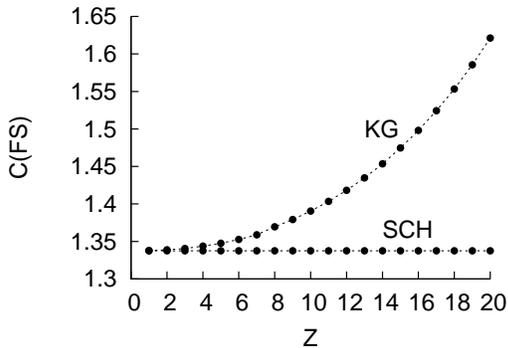}
\end{center}
\caption{Fisher-Shannon complexity for the ground state Klein-Gordon (KG) and Schr\"{o}dinger (SCH) pionic atom in terms 
of the nuclear charge $Z$ (atomic units are used).}
\end{figure}

To measure the relativistic effects we define the quantity $\zeta_{\textrm{FS}}=1-\frac{C_{SCH}(FS)}{C_{KG}(FS)}$. This 
quantity varies from $0$ to $1$, so that $\zeta_{\textrm{FS}}\sim0$  when the relativistic effects are negligible and  
$\zeta_{\textrm{FS}} \sim 1$ in the ultrarelativistic limit. In Figure 2 we can see the effects of the relativity model 
for different values of the nuclear charge $Z$ and various $S(n,0,0)$ states.
First, we observe that the relativistic effects increase when the nuclear charge is increasing not only 
for the ground state (as already pointed out) but also for all the excited states. Second, the relativistic
effects decrease when the principal quantum number is increasing. Third, this decreasing behaviour 
with $n$ has a strong dependence with $Z$, being slower as bigger is the nuclear charge.

\begin{figure}[h]
\begin{center}
\includegraphics[scale=0.45]{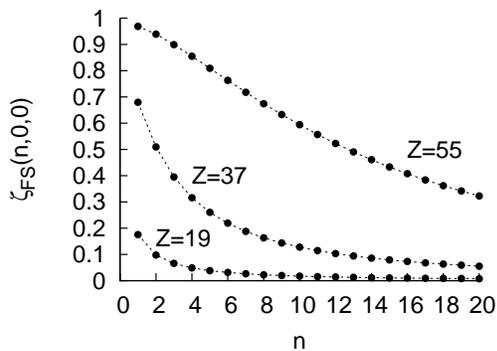}
\end{center}
\caption{Relative ratio of the Fisher-Shannon complexity for various $S$ states of the Klein-Gordon and Schr\"{o}dinger pionic atom.}
\end{figure}

\begin{figure}[h]
\begin{center}
\includegraphics[angle=270, scale=0.18]{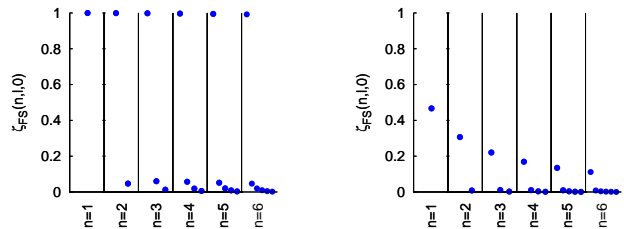}
\end{center}
\caption{Relative ratio of the Fisher-Shannon complexity for various states $(n,l,0)$ of the Klein-Gordon and Schr\"{o}dinger pionic atoms 
with $Z=19$ (left) and $Z=55$ (right).}
\end{figure}

In Figure 3 we can observe that the relativistic effects are practically negligible when the angular quantum number is 
different to cero even when the nuclear charge is big. This dependence in $l$ is more important than the dependence 
on the principal quantum number $n$. For completeness, let us point out that the relativistic effects are practically 
negligible when the magnetic quantum number $m$ varies for $(Z,n,l)$ fixed.

In conclusion, we have explored relativistic effects on the behaviour of the Fisher-Shannon complexity of
pionic systems with nuclear charge $Z$ in the Klein-Gordon framework. We have done it for both ground and excited states.
First we found that the relativistic Fisher-Shannon complexity grows when the nuclear charge increases in contrast with 
the non-relativistic case for both ground and excited states. A similar behaviour has been recently observed in the  
case of the ground state of systems governed by the Dirac equation \cite{B07,B08,S08,S09}. We found that this trend 
remains for excited states in a damped way, so that the relativistic effects enhance with $Z$ for a given $(n,l,m)$ 
state and, for a given $Z$,  decrease when the principal and/or orbital quantum numbers  are increasing. Let us also highlight that the 
non-relativistic limits at large principal quantum number $n$ for a given $Z$ (see Figs. 3 and 4) and at small valuesof $Z$ (see Fig. 2) are reached. 
On the other hand it is pertinent to underline that the finite nuclear volume effects are very tiny for any 
information-theoretic and complexity measure because of its macroscopic character.

It is worthwhile noting that the Fisher-Shannon complexity given by
Eq. (\ref{eq:fs}) is a much better qualitative and quantitative measure of
relativistic effects than any other single information-theoretic
measure, including the Shannon entropy and Fisher information recently
studied \cite{M08}, because it grasps both the total extent and the gradient
content of the quantum-mechanical probability densities which
characterize the hydrogenic states. Moreover, contrary to the Shannon and
Fisher quantities, the Fisher-Shannon complexity better reflects the usual
understanding of complexity of the system because of its special
properties; namely, minimization at the extreme ordered and disordered
limits and invariance under replication, translation and rescaling
transformations

Let us finally say for completeness that we have also investigated the relativistic Klein-Gordon effects in pionic atoms 
by means of the LMC shape complexity \cite{L95}  $C(\textrm{LMC})=\left < \rho \right > \exp{S\left[ \rho \right]}$. We 
found that the relativistic effects are also identified by this quantity but in a much weaker way than the Fisher-Shannon 
complexity $C(\textrm{FS})$. Apparently this is because of the property of locality of $C_{\textrm{FS}}$ coming through its 
gradient-dependent Fisher ingredient, which graps much better the (strong) oscillatory condition of the pionic densities.

The authors gratefully acknowledge the Spanish MICINN grant FIS2008-02380 and the grants FQM-2445 and FQM-4643 of the Junta 
de Andaluc\'ia. They belong to the research group FQM-207. S.L.R. and D.M. acknowledge the FPU and FPI scholarships of the 
Spanish Ministerio de Ciencia e Innovaci\'{o}n, respectively. The authors want to aknowledge B. Janssen and A.R. Plastino
for useful conversations.

\end{document}